\documentclass[%
 reprint,
superscriptaddress,
 amsmath,amssymb,
 aps,
 twocolumn,
 pre,
floatfix,
]{revtex4-1}

\usepackage{graphicx}
\usepackage{dcolumn}
\usepackage{bm}
\usepackage{xcolor}
\usepackage[caption=false]{subfig}
\usepackage{hyperref}

\usepackage{cleveref}
\crefformat{section}{\S#2#1#3}

\begin{document}

\preprint{APS/123-QED}

\title{Premelting controlled active matter in ice}

\author{J\' er\' emy Vachier}
\email{jeremy.vachier@su.se}
\affiliation{Nordita, KTH Royal Institute of Technology and Stockholm University, Hannes Alfv\' ens v\" ag 12, SE-106 91 Stockholm, Sweden}
\author{J. S. Wettlaufer}
\email{john.wettlaufer@yale.edu}
\affiliation{Nordita, KTH Royal Institute of Technology and Stockholm University, Hannes Alfv\' ens v\" ag 12, SE-106 91 Stockholm, Sweden}
\affiliation{%
 Yale University, New Haven, Connecticut 06520-8109, USA \\
 }%

\date{\today}

\begin{abstract}
Self-propelled particles can undergo complex dynamics due to a range of bulk and surface interactions. When a particle is embedded in a host solid near its bulk melting temperature, the
latter may melt at the surface of the former in a process known as interfacial premelting. The thickness of the melt film depends on the temperature, impurities, material properties and geometry. A temperature gradient is accompanied by a thermomolecular pressure gradient that drives the interfacial liquid from high to low temperatures and hence the particle from low to high temperatures, in a process called thermal regelation. When the host material is ice and the embedded particle is a biological entity, one has a particularly novel form of active matter, which addresses interplay between a wide range of problems, from extremophiles of both terrestrial and exobiological relevance to ecological dynamics in Earth's cryosphere. Of basic importance in all such settings is the combined influence of biological activity and thermal regelation in controlling the redistribution of bioparticles.  
Therefore, we re-cast this class of regelation phenomena in the stochastic framework of active Ornstein-Uhlenbeck dynamics and make predictions relevant to this and related problems of interest in biological and geophysical problems. 
We examine how thermal regelation compromises paleoclimate studies in the context of ice core dating and we find that the activity influences particle dynamics during thermal regelation by enhancing the effective diffusion coefficient.  Therefore, accurate dating relies on a quantitative treatment of both effects.
\end{abstract}

\maketitle


\section{\label{sec:intro}Introduction}
Glaciers, ice sheets, sea ice and permafrost constitute
large ecosystems and cover significant areas of the planet
\cite{anesio-nature-2017,pandey-science-2016,abramov-review-2021,margesin-book-2008}. 
Ice cores provide the highest resolution records of paleoclimate for the last eight glacial cycles \cite{stauffer-ag-2004,royer-springer-1983}. Therefore, ice core dating methods are of primary importance in quantifying climate processes and providing empirical constraints for models used to predict future climates. 
Moreover, the cryosphere is rich with biological activity, from algae, such as \textit{Chlamydomonas nivalis} or \textit{Euglena viridis}, to diatoms \cite{anesio-nature-2017,abramov-review-2021, wager-jstor-1911} and bacteria, such as \textit{Pseudomonas priestleyi} or \textit{Pseudomonas syringae} \cite{anesio-nature-2017,pandey-science-2016}. 
Through physical and/or chemical interactions, these micro-organisms interact with their surroundings thereby creating environmental feedback \cite{anesio-nature-2017,pandey-science-2016}. An important property of most of these micro-organisms is their motility \cite{gompper-iop-2020,bechinger-rmp-2016,margesin-book-2008} and they commonly evolve and hence self-propel adjacent to ice-water interfaces. 
For example, \textit{Williamson et al.} \cite{williamson-pnas-2020} and \textit{Ryan et al.} \cite{ryan-nature-2018} have recently shown that biota can accelerate the melting of ice in Greenland. Moreover, studies of sea ice support the idea that the migration of a range of bacteria,
eukaryotes and prokaryotes within the ice column is accompanied by a phase change. As a consequence, a symbiotic
relationship between sea ice biota and phase change is
observed \cite{van-esa-2018,aumack-jms-2014,lindensmith-plos-2016}. In contrast, the bacteria \textit{Pseudomonas
syringae} acts as a heterogeneous ice nucleus, and is used
to create artificial snow \cite{pandey-science-2016}. 
Some of these of living micro-organisms persist in ice and permafrost for centuries. For example, recently a 30,000 year old giant virus {\textit{Pithovirus sibericum}} has been found in permafrost \cite{legendre-pnas-2014}, along with microbes \cite{zhong-micro-2021,el-espr-2021} and nematodes \cite{shatilovich-book-2018}. Moreover, viable bacteria have been found in 750,000 year old glacial ice \cite{christner-em-2003}. Finally, not only can microscopic species survive in subfreezing conditions, but the clams \textit{Arctica islandica} can live in low temperature extremes for up to 400 years \cite{ungvari-jos-2011,philipp-g-2010}. 
Finally, the setting we study is of importance in understanding extremophiles.  
For example, it has been shown that the evolution of microbial cells can be linked to permafrost age \cite{abramov-review-2021}, a {\em living} Bdelloid rotifer has recently been recovered from permafrost 24,000 years old \cite{shmakova-cell-2021}, and a new species has been discovered in 16 million-year-old amber \cite{Mapalo:2021}.
Therefore, understanding the physico-chemical relationship between microorganisms and their environment underlies key questions concerning the covariation of life and climate.\newline
\indent Active particles are able to convert energy from biological, chemical, or physical processes into motion, and
can exhibit macroscopic behavior that  can lead to the
emergence of collective motion \citep[e.g.,][]{elgeti-rpp-2015,romanczuk-epjst-2012}. Due to the wide
range of implications of their dynamics and their role as
a model system in non-equilibrium statistical mechanics, active particles have been the focus of a great deal
of attention in the past several decades \citep[e.g.,][]{cates-rpp-2012,bechinger-rmp-2016,ghosh-nl-2009,kim-am-2013,jin-pre-2019}. Generally, active particles evolve in an aqueous environment where, because of their microscopic size, viscous forces dominate over inertial
forces. Biological active particles, such as algae and bacteria, operate
in complex geometries, for example in membranes, glaciers
and ice sheets, and can react to chemical, physical and biological gradients, most commonly associated with nutrients and waste \cite{madigan-book-2008,adler-science-1966,hazelbauer-arm-2012}. In the dilute case, where the
particle-particle interactions can be neglected, the dynamics is dominated by the balance of active motion and
external forces, such as gravity \cite{vachier-epje-2019,palacci-prl-2010,ginot-npj-2018,hermann-sm-2018} or temperature gradients.

When a particle is embedded in ice near its bulk melting temperature, the ice may melt at the particle-ice surface in a process known as \textit{interfacial premelting}. The thickness of the melt film depends on temperature, impurities, material properties and geometry \cite{dash-rmp-2006physics}. A temperature gradient is accompanied by a thermomolecular pressure gradient that drives the interfacial liquid from high to low temperatures and hence the particle from low to high temperatures, in a process called \textit{thermal regelation} \cite{peppin-jsp-2009,dash-rmp-2006physics,wettlaufer-arfm-2006,rempel-prl-2001,marath-sm-2020}.

\begin{figure}[t]
\centering
\includegraphics[scale=.53]{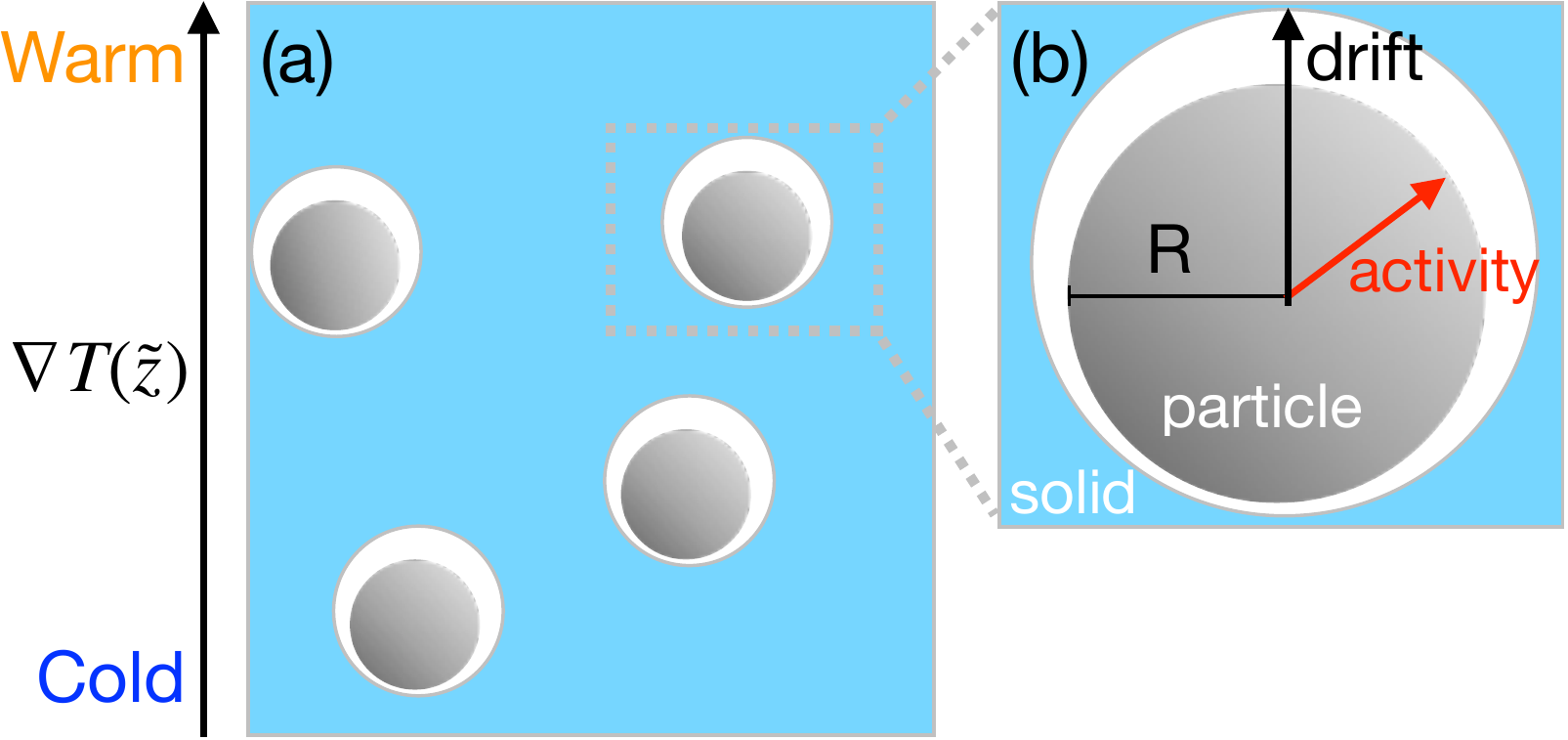}
\caption{(a) Perspective view of few active particles embedded inside a solid against which they premelt and experience an external temperature gradient $\nabla T$. (b) An expanded view of one active particle inside the solid. The radius of the particle is $R$, the black arrow shows the drift velocity induced by thermal regelation and the red arrow denotes the activity given by an active force.}
\label{figopening}
\end{figure}

Thermal regelation of inert particles plays a major role in the redistribution of material inside of ice, which has important environmental and composite materials implications \cite{peppin-jsp-2009,dash-rmp-2006physics,marath-sm-2020,wettlaufer-arfm-2006,rempel-prl-2001,worster-jfm-2021}. However, surface properties are not only central to the physical processes shaping Arctic \cite{Hallet:2013} and Antarctic \cite{Liu:2018} landscapes, but they underlie the fact that extremophile organisms on Earth develop strategies to act back on their harsh environments in order to improve living conditions.
Important examples include exopolymeric substances and antifreeze glycoproteins, both of which have unique impacts on the surface of ice that enhance liquidity \cite{hansen-pss-2014}. The confluence of thermal
regelation, bio-enhanced premelting and intrinsic mobility motivate the work described here.

The case of active particles in premelting solids is illustrated in Fig. \ref{figopening}.  The fact that intrinsic mobility, defined by an active force, may compete with thermal regelation, induced by a temperature gradient, has not been previously studied, and is particularly relevant for biota in ice. For example, one known effect of the activity on a particle's dynamics is to change its diffusion properties  \cite{bechinger-rmp-2016}. Treating this question is particularly relevant for ice core paleoclimate dating methods due to the long periods of time over which diffusion can act \cite{royer-springer-1983,rupper-jg-2015,baccolo-tcd-2021,augustin-nature-2004}. Additionally, it has been shown in \cite{speck-epl-2016,puglisi-mdpi-2017,marconi-sc-2017,schmidt-prl-2018,kummel-sm-2015} that active particles dissipate energy locally in the system via the active force. However, quantifying this effect experimentally is still very challenging due to the wide range of other biological mechanisms at play.

The framework of our study is that of a so-called active Ornstein-Uhlenbeck particle (AOUP) \citep[e.g][]{bonilla-pre-2019,fodor-prl-2016,caprini-sm-2018,martin-pre-2021,dabelow-fp-2021,caprini-sr-2019,dabelow-prx-2019}.
The AOUP is described by an active force subject to a stochastic evolution that can be modeled by an Ornstein-Uhlenbeck process. This process is characterized by a persistence time, $\tau_a$, which is the time scale after which the system switches from a ballistic to a diffusive regime, and a variance given by $\tilde{D}_a/\tau_a$.  A consequence of this is that the diffusivity is increased and the resulting diffusion coefficient is $\tilde{D}_a$, which can be compared to a colored noise process \cite{martin-pre-2021,sevilla-pre-2019}.
Both of these two key parameters, $\tau_a$ and $\tilde{D}_a$, can be measured experimentally \cite{maggi-prl-2014,maggi-sr-2017,donado-sr-2017,wu-aem-2006}. 

The AOUP describes the motion of colloids in a bath
of active particles \cite{bechinger-rmp-2016,maggi-prl-2014,maggi-sr-2017} and, in the case of a bacterial
bath, \textit{Wu and Libchaber} \cite{wu-prl-2000} showed that the activity of the bacteria enhanced the diffusion of passive tracer particles by two to three orders of magnitude relative to the thermal case. Finally, while the AOUP model provides
accurate predictions for a range of complex phenomena \cite{caprini-sm-2018,dabelow-fp-2021,caprini-sr-2019,marconi-sm-2016}, in contrast to the so-called Active Brownian Particle (ABP) and Run-and-Tumble models, a theoretical advantage of the AOUP is its Gaussian nature \cite{martin-pre-2021}. Moreover, similar to the ABP,
the long term behavior of the AOUP is also diffusive \cite{caprini-sr-2019}. These issues motivate our use of the AOUP model framework to describe the motion of active particles in ice under an external temperature gradient. We analyze these particles in three dimensions using a multiple scale expansion to derive the associated Fokker-Planck equation. Similar approaches have been used in the case of a passive Brownian particle \cite{aurell-epl-2016,aurell-pre-2017,bo-jsm-2019} and for an active Brownian particle in a channel \cite{chen-arxiv-2020}.\newline
\indent Our paper is organized as follows. In \cref{sec:method} we outline the active Ornstein-Uhlenbeck particle model. In \cref{sec:results} we derive the associated Fokker-Planck equation using a multiple scale expansion and then find the analytic solution in the limit that regelation dominates the dynamics.  
We then compare our analytic solutions with numerical solutions before concluding in \cref{sec:conclusion}.

\section{\label{sec:method}Method}
As shown in Fig. \ref{figopening}, due to thermal regelation, the motion of the active particle is biased by a drift velocity $\tilde{v}(\tilde{z})=U(\tilde{z})\hat{\tilde{\bm{z}}}$ in the direction of the temperature gradient \cite{marath-sm-2020}. Regelation is a consequence of the premelted film around the particle, which can also execute diffusive motion in the ice column. The premelting-controlled diffusivity of the particle is given by $\tilde{D}(\tilde{z})\mathbb{I}$, where $\mathbb{I}$ is the identity matrix \cite{marath-sm-2020}. 
The evolution of the particle's position, $\tilde{\bm{r}}=(\tilde{r}_1,\tilde{r}_2,\tilde{r}_3)=(\tilde{x},\tilde{y},\tilde{z})$, and its activity are described by two overdamped Langevin equations
\begin{align}
\label{eq:lange_posi}
\frac{d}{d\tilde{t}}\tilde{\bm{r}}(\tilde{t}) &= \sqrt{2\tilde{D}_a}\tilde{\bm{\eta}}(\tilde{t})+ \tilde{v}(\tilde{z}) +\sqrt{2\tilde{D}(\tilde{z})}\bm{\xi}_p(\tilde{t}) \,,\\
\label{eq:lange_eta}
\frac{d}{d\tilde{t}}\tilde{\bm{\eta}}(\tilde{t}) &= -\frac{1}{\tau_a}\tilde{\bm{\eta}}(\tilde{t}) + \frac{1}{\tau_a}\bm{\xi}_a(\tilde{t}) \,.
\end{align}
The activity, or self-propulsion, is given by the term $\sqrt{2\tilde{D}_a}\tilde{\bm{\eta}}$ in Eq. \eqref{eq:lange_posi}, where $\tilde{D}_a$ is the active diffusivity representing the active fluctuations of the system, which result from the interactions between tracer particles and their surrounding environment. 
Interesting examples include, among others, coupling to a viscoelastic medium, such as a cytoskeleton, or a bacterial bath \cite{vandebroek-sm-2017,romanczuk-prl-2011,joanny-prl-2003,peruani-prl-2007}.

The function $\tilde{\bm{\eta}}=(\tilde{\eta}_1,\tilde{\eta}_2,\tilde{\eta}_3)$ is described by an Ornstein-Uhlenbeck process, with correlations given by
\begin{equation}
\langle\tilde{\eta_i}(\tilde{t}')\tilde{\eta_j}(\tilde{t}) \rangle = \frac{\delta_{ij}}{\tau_a}e^{-\frac{|\tilde{t}'-\tilde{t}|}{\tau_a}}\,,
\label{eq:OU}
\end{equation}
where $\tau_a$ is the noise persistence. In the small $\tau_a$ limit, $\tilde{\bm{\eta}}$ reduces to  Gaussian white noise with correlations $\langle \tilde{\eta}_i(t')\tilde{\eta}_j(t)\rangle = \delta_{ij}\delta(\tilde{t}'-\tilde{t})$. In contrast, when $\tau_a$ is finite, $\tilde{\bm{\eta}}$ does not reduce to Gaussian white noise, and Eq. \eqref{eq:lange_posi} does not reach equilibrium. Hence, $\tau_a$ controls the non-equilibrium properties of the system \cite{martin-pre-2021,caprini-sr-2019,caprini-mpi-2021}.  
The random fluctuations are given by zero mean Gaussian white noise processes $\langle \xi_{p_i}(\tilde{t}')\xi_{p_j}(\tilde{t}) \rangle = \delta_{ij}\delta(\tilde{t}'-\tilde{t})$ and $\langle \xi_{a_i}(\tilde{t}')\xi_{a_j}(\tilde{t}) \rangle = \delta_{ij}\delta(\tilde{t}'-\tilde{t})$. 
In the limit where soluble impurities control the premelted film
thickness in ice, the velocity and diffusivity are given by 
\begin{equation}
U(\tilde{z}) = - \frac{A_3}{A_2^3}\frac{1}{\tilde{z}^3}\,,
\label{eq:drift}
\end{equation}
and
\begin{equation}
\tilde{D}(\tilde{z}) = \frac{(R_gT_mN_i)^3}{8\pi \nu R^4 A_2^3}\frac{k_BT_m}{\tilde{z}^3}\,,
\label{eq:diff}
\end{equation}
with $A_2= \rho_lq_m\frac{|\nabla T|}{T_m}$ and $A_3 = \rho_sq_m|\nabla T|\frac{(R_gT_mN_i)^3}{6\nu RT_m}$ \cite{marath-sm-2020}. The universal gas constant is $R_g$;  the latent heat of fusion per mole of the solid is $q_m$;  the molar density  of the liquid is $\rho_l$; the magnitude of the external temperature gradient is $|\nabla T|$; the pure bulk melting temperature is $T_m= 273.15$K; the viscosity of the fluid is $\nu$; the particle radius is $R$; the number of moles of impurities per unit area of the interface is $N_i$; $k_B$ is Boltzmann's constant; and $\rho_sq_m\sim 334 \times 10^{6}$ J$\cdot$m$^{-3}$ \cite{marath-sm-2020}.

The Langevin Eqs. \eqref{eq:lange_posi} and \eqref{eq:lange_eta}, allow us to express the probability of finding a particle at the position $\tilde{\bm{r}}=(\tilde{r}_1,\tilde{r}_2,\tilde{r}_3)=(\tilde{x},\tilde{y},\tilde{z})$ at a given time $t$ through the Fokker-Planck equation, which describes the evolution of
the probability density function $P(\tilde{\bm{r}},\tilde{\bm{\eta}},\tilde{t}|\tilde{\bm{r}}_0,\tilde{\bm{\eta}}_0,\tilde{t}_0)$, with the initial condition $P(\tilde{\bm{r}},\tilde{\bm{\eta}},\tilde{t}=\tilde{t}_0|\tilde{\bm{r}}_0,\tilde{\bm{\eta}}_0,\tilde{t}_0)=\delta(\tilde{\bm{r}}-\tilde{\bm{r}}_0)\delta(\tilde{\bm{\eta}}-\tilde{\bm{\eta}}_0)$. To simplify the notation, we write the conditional probability as $P(\tilde{\bm{r}},\tilde{\bm{\eta}},\tilde{t})=P(\tilde{\bm{r}},\tilde{\bm{\eta}},\tilde{t}|\tilde{\bm{r}}_0,\tilde{\bm{\eta}}_0,\tilde{t}_0)$ and eventually arrive at the following Fokker-Planck equation \cite{fodor-prl-2016,martin-pre-2021}
{\small
\begin{align}
\frac{\partial}{\partial \tilde{t}}  P(\tilde{\bm{r}},\tilde{\bm{\eta}},\tilde{t})&=-\frac{\partial}{\partial \tilde{r}_3}\left[ \tilde{v}(\tilde{r}_3) P(\tilde{\bm{r}},\tilde{\bm{\eta}},\tilde{t})\right]-\sqrt{2\tilde{D}_a}\tilde{\bm{\eta}}\cdot\nabla_{\tilde{\bm{r}}} P(\tilde{\bm{r}},\tilde{\bm{\eta}},\tilde{t})\nonumber\\
&+\nabla^2_{\tilde{\bm{r}}}\left[\tilde{D}(\tilde{r}_3) P(\tilde{\bm{r}},\tilde{\bm{\eta}},\tilde{t})\right]+\frac{1}{\tau_a}\nabla_{\tilde{\bm{\eta}}}\cdot\left[\tilde{\bm{\eta}} P(\tilde{\bm{r}},\tilde{\bm{\eta}},\tilde{t}) \right]\nonumber\\
&+\frac{1}{2\tau_a^2}\nabla^2_{\tilde{\bm{\eta}}} P(\tilde{\bm{r}},\tilde{\bm{\eta}},\tilde{t})\,.
\label{eq:fpe3D}
\end{align}}
Although Eq. \eqref{eq:fpe3D} contains both microscopic and
macroscopic scales, we are interested in the long term
behavior and hence seek to extract the effective “macroscopic dynamics” from it, which we describe presently.

\section{\label{sec:results}Results}
\subsection{\label{subsec:multiscale}Method of multiple scales}
The macroscopic length characterizing the heat flux is
\begin{equation}
L=\frac{T_m}{|\nabla T|}\,.
\label{eq:macrolength}
\end{equation}
The particle scale $l$ is such that $l << L$, and hence we can define a small parameter $\epsilon$ as
\begin{equation}
\epsilon = \frac{l}{L}\,.
\label{eq:epsilon}
\end{equation}
We introduce the following dimensionless variables
\begin{align}
&\bm{\eta}= \sqrt{\tau_a}\tilde{\bm{\eta}}\text{, } \bm{r} = \frac{\tilde{\bm{r}}}{l} \text{, } t =\frac{\tilde{t}}{\tau} \text{, }v=\frac{\tilde{v}}{u} \text{, } v_a = \frac{\tilde{v}_a}{v_{ac}}\nonumber\\
& \text{ and }D = \frac{\tilde{D}}{D_c} \,,
\label{eq:dimensionless_quanti}
\end{align}
where $\tilde{v}_a = \sqrt{\frac{2\tilde{D}_a}{\tau_a}}$ \cite{dabelow-jsm-2021}, 
$v_{ac}$ is the characteristic active velocity, and $u$ and $D_c$ are the characteristic regelation speed and premelting controlled diffusivity respectively. In Eq.\eqref{eq:dimensionless_quanti}, we non-dimensionalized the space variable, $\tilde{\bm{r}}$, with the microscopic length, $l$, and the time variable, $\tilde{t}$, with a characteristic time, $\tau$, the latter to be determined \textit{a posteriori}.
Thus Eq. \eqref{eq:fpe3D} becomes
\begin{align}
P_l\frac{\partial}{\partial t} P &= -P_{e}\frac{\partial}{\partial r_3}\left[ vP\right]-P_{a}v_a\eta\nabla_{\bm{r}}P+\nabla^2_{\bm{r}}\left[DP \right]\nonumber\\
&+P_{A}\nabla_{\bm{\eta}}\left[\bm{\eta}P \right]+\frac{1}{2}P_{A}\nabla^2_{\bm{\eta}}P\,,
\label{eq:fpe_dimless}
\end{align}
in which the following dimensionless numbers appear;
\begin{equation}
P_{e}=\frac{ul}{D_c} \text{, } P_{a}=\frac{v_{ac}l}{D_c} \text{, } P_l=\frac{l^2}{D_c\tau} \text{ and } P_{A}=\frac{l^2}{D_c\tau_a}\,.
\end{equation}
The first two numbers $P_e$ and $P_a$ are the P\' eclet numbers associated with regelation and activity respectively.
We identify four characteristic time scales: 
$t_l^{\text{diff}}=l^2/D_c$, $t_l^{\text{adv}}=l/u$, $t_L^{\text{diff}}=L^2/D_c$ and $t_L^{\text{adv}}=L/u$    
associated with microscopic (macroscopic) diffusion and advection on the
particle scale $l$ (thermal length scale $L$). 
The ratio of characteristic time for diffusion and advection is the P\' eclet number
\begin{equation}
P_{e} = \frac{t_l^{\text{diff}}}{t_l^{\text{adv}}} \text{ and } P_{e}^L = \frac{t_L^{\text{diff}}}{t_L^{\text{adv}}}\,.
\end{equation}
As a result of the external temperature gradient, our system is driven by thermal regelation, therefore advection dominates on the macroscopic scale and $P_e^L=\mathcal{O}(1/\epsilon)$, or equivalently, $t_L^{\text{adv}}=\epsilon t_L^{\text{diff}}$. However, because $P_{e}=\mathcal{O}(1)$ we have $t_l^{\text{adv}}= t_l^{\text{diff}}$.  Thus, on the macroscopic scale, the P\' eclet number $P_e^L$ is large, advection dominates, and we use the macroscopic advection time $\tau=t_L^{\text{adv}}$ as our characteristic time, leading to 
\begin{align}
\epsilon\frac{\partial}{\partial t} P &= -\frac{\partial}{\partial r_3}\left[ vP\right]-P_{a}v_a\eta\nabla_{\bm{r}}P+\nabla^2_{\bm{r}}\left[DP \right]\nonumber\\
&+P_{A}\nabla_{\bm{\eta}}\left[\bm{\eta}P \right]+\frac{1}{2}P_{A}\nabla^2_{\bm{\eta}}P\,. 
\label{eq:fpe_dimless_scenaadv}
\end{align}
We introduce a dimensionless macroscopic length, $\bm{R}=\tilde{\bm{r}}/L$, and a dimensionless microscopic time, $T=\tilde{t}/t^{\text{adv}}_l$, thereby stretching the scales
\begin{equation}
\bm{r}=\frac{1}{\epsilon}\bm{R}\text{ and }T = \frac{1}{\epsilon}t\,.
\label{eq:sep_scale}
\end{equation}
Now, invoking a power series ansatz 
\begin{equation}
P = P^0 + \epsilon P^1 + \epsilon^2 P^2\,,
\label{eq:p_expansion}
\end{equation}
we derive a system of equations at each order in $\epsilon$ \cite{bensoussan-book-2011,sanchez-book-1980,bender-book-2013}, which are
\begin{widetext}
\begin{align}
\mathcal{O}(\epsilon^0): \mathcal{L}P^0 &= 0\,,\label{eq:hiere_1}\\
\mathcal{O}(\epsilon):\mathcal{L}P^1 &= \frac{\partial}{\partial T}P^0+\frac{\partial}{\partial R_3}\left[ vP^0\right]+P_{a}v_a\bm{\eta}\cdot\nabla_{\bm{R}} P^0-2\nabla_{\bm{r}}\cdot\nabla_{\bm{R}}\left[DP^0\right]\,,\label{eq:hiere_2}\\
\mathcal{O}(\epsilon^2):\mathcal{L}P^2 &= \frac{\partial}{\partial T}P^1+\frac{\partial}{\partial t}P^0+\frac{\partial}{\partial R_3}\left[ vP^1\right] +P_{a}v_a\bm{\eta}\cdot\nabla_{\bm{R}}P^1-2\nabla_{\bm{r}}\cdot\nabla_{\bm{R}}\left[ DP^1\right]\nonumber\\
&-\nabla_{\bm{R}}^2\left[D P^0\right]\label{eq:hiere_3}\,,
\end{align}
\end{widetext}
where $\mathcal{L} = \mathcal{M}+\mathcal{Q}$, with $\mathcal{M} = -\frac{\partial}{\partial r_3}v-P_{a}v_a\bm{\eta}\cdot\nabla_{\bm{r}}+\nabla^2_{\bm{r}}D$, and $\mathcal{Q} = P_{A}\nabla_{\bm{\eta}}\cdot\bm{\eta}+\frac{P_{A}}{2}\nabla^2_{\bm{\eta}}$. Following \cite{pavliotis-book-2008,aurell-epl-2016} we solve  Eqs. \eqref{eq:hiere_1}-\eqref{eq:hiere_3}. The solution of the leading order Eq. \eqref{eq:hiere_1} is derived by making the following product ansatz
\begin{equation}
P^0(\bm{r},\bm{R},\bm{\eta},T,t)=w(\bm{r},\bm{\eta})\rho^0(\bm{R},T,t)\,.
\label{eq:1ansatz}
\end{equation} 
We integrate by parts over the microscale variables $\bm{r}$ and $\bm{\eta}$, and use periodic boundary conditions to obtain the so-called weak formulation of the leading order equation \cite{auriault-book-2010,Chipot2009}. The existence and uniqueness of $P^0$ is insured by use of the Lax-Milgram theorem \cite{auriault-book-2010,hsiao-book-2008}, otherwise known as the solvability condition \cite{jikov-book-2012,pavliotis-book-2008,bakhvalov-book-2012} or the Fredholm alternative  \cite{kress-jde-1977,reed-book-2012}. 
Thus, $P^0$ is constant over $P^0(\bm{r},\bm{R},\bm{\eta},T,t)=P^0(\bm{R},\bm{\eta},T,t)$ and Eq. \eqref{eq:hiere_1} becomes
\begin{equation}
\nabla_{\bm{\eta}}\cdot\left[\bm{\eta} w\right]+\frac{1}{2}\nabla^2_{\bm{\eta}}w = 0\,.
\end{equation}
Using the known result for a multi-dimensional Ornstein-Uhlenbeck process \cite{risken-book-1996}, the solution for $w$ is given by
\begin{equation}
w(\eta_1,\eta_2,\eta_3) = \prod\limits_{i=1}^3 \frac{1}{\sqrt{2\pi}}e^{\dfrac{-\eta^2_i}{2}}\,.
\label{eq:w_solution}
\end{equation}
The solvability condition for the equation at $\mathcal{O}(\epsilon)$ is 
\begin{align}
\int d\bm{r}d\bm{\eta} &\left( w\frac{\partial}{\partial T}\rho^0+w\frac{\partial}{\partial R_3}\left[v\rho^0\right]+wP_av_a\bm{\eta}\cdot\nabla_{\bm{R}}\rho^0\right) \nonumber\\
&= 0\,,
\label{eq:solvability_seond}
\end{align}  
which depends on the leading order result, $P^0$, from which we find that 
\begin{equation}
\frac{\partial}{\partial T}\rho^0 = -\frac{\partial}{\partial R_3}\left[v \rho^0 \right]\,,
\label{eq:solvability_second_step1}
\end{equation}
and Eq. \eqref{eq:hiere_2} becomes
\begin{equation}
\mathcal{L}P^1 = wP_av_a\bm{\eta}\cdot\nabla_{\bm{R}}\rho^0\,.
\label{eq:solvability_second_step2}
\end{equation}
From Eq. (24), we observe that $P^1$ depends linearly on the first derivative, $\nabla_{\bm{R}}\rho^0$, and using separation of variables we can write the solution as
\begin{equation}
P^1=wP_av_a\bm{\alpha}\cdot\nabla_{\bm{R}}\rho^0\,,
\label{eq:2ansatz}
\end{equation}
where $\bm{\alpha}$ is the so-called first order corrector \cite[e.g.,][]{pavliotis-book-2008}, from which we find that
\begin{equation}
\bm{\alpha} = -\frac{1}{P_{A}}\bm{\eta}\,.
\label{eq:solution_1auxiliary}
\end{equation}
Using $P^1$ and the solvability condition in Eq. \eqref{eq:hiere_3} we obtain
\begin{equation}
\frac{\partial}{\partial t}\rho^0 = \frac{P_a^2v_a^2}{2P_{A}}\nabla_{\bm{R}}^2\rho^0+\nabla_{\bm{R}}^2\left[D\rho^0 \right]\,,
\label{eq:fpe_dimensionless}
\end{equation}
which in dimensional form is
\begin{equation}
\frac{\partial}{\partial\tilde{t}}\rho=-\frac{\partial}{\partial\tilde{r}_3}\left[\tilde{v}\rho \right]+\tilde{D}_a\nabla_{\tilde{\bm{r}}}^2\rho+\nabla_{\tilde{\bm{r}}}^2\left[\tilde{D}(\tilde{r}_3)\rho \right]\,.
\label{eq:effective_fpe}
\end{equation}
This Fokker-Planck equation is valid to leading order, $P^0$, viz., Eq. (15), and captures the behavior on long time scales.  
At this order the active force is embodied in the effective diffusivity, leading to an increase in the diffusivity induced by thermal regelation. This picture is consistent with previous work in active systems \cite{vachier-epje-2019,bechinger-rmp-2016,cugliandolo-fnl-2019,caprini-sm-2018}.  However higher order contributions \cite[c.f.,][]{mauri-pefa-1991} are beyond the scope of this study.

\subsection{\label{subsec:largepeclet}Solution of the Fokker-Planck equation in the large P\' eclet function limit}

\begin{figure}
    \subfloat[]{%
  	\includegraphics[width=1.\columnwidth]{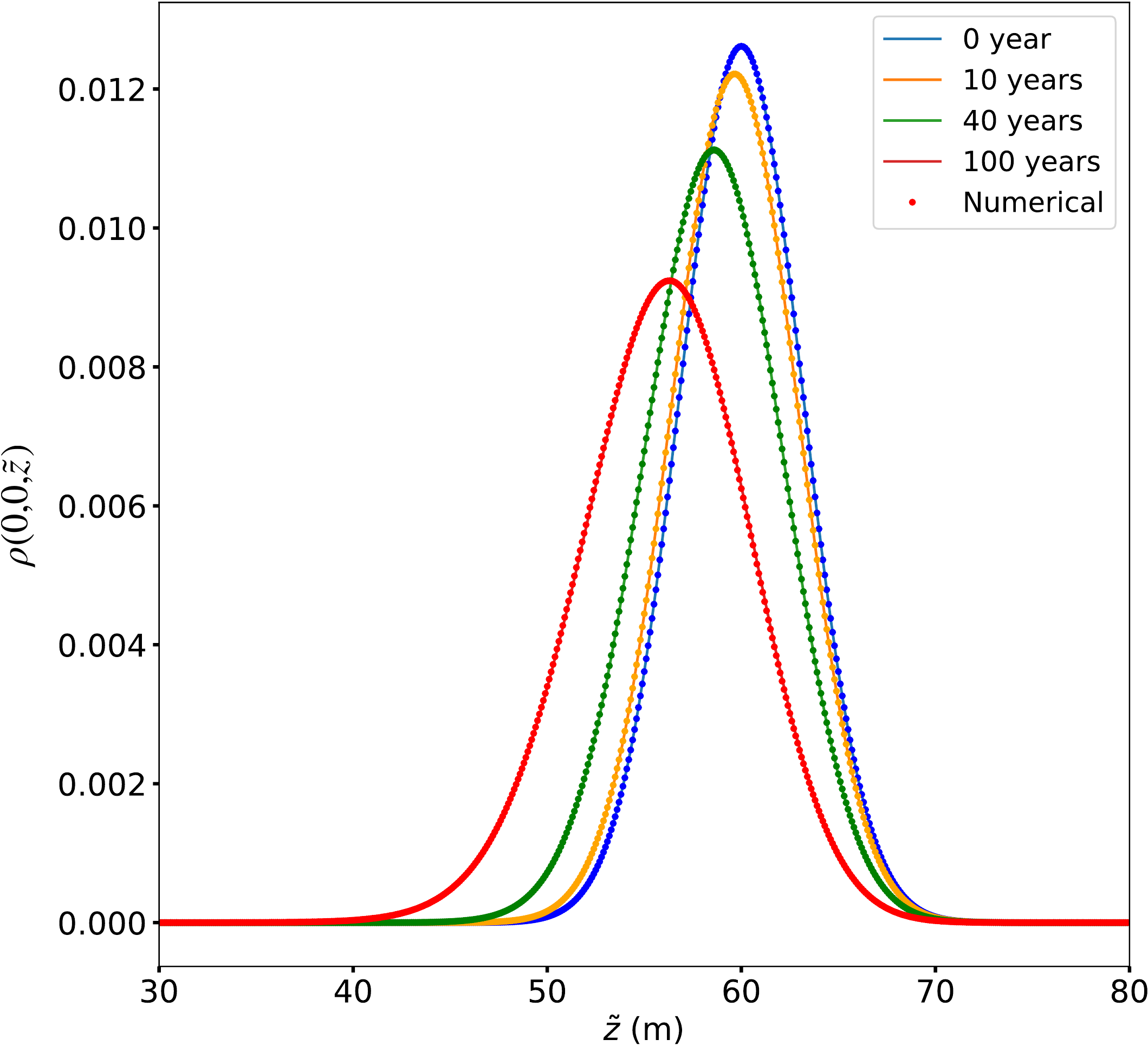}%
\label{fig1a:passivecompaz}}\hfill
	 
	\subfloat[]{%
  	\includegraphics[width=1.\columnwidth]{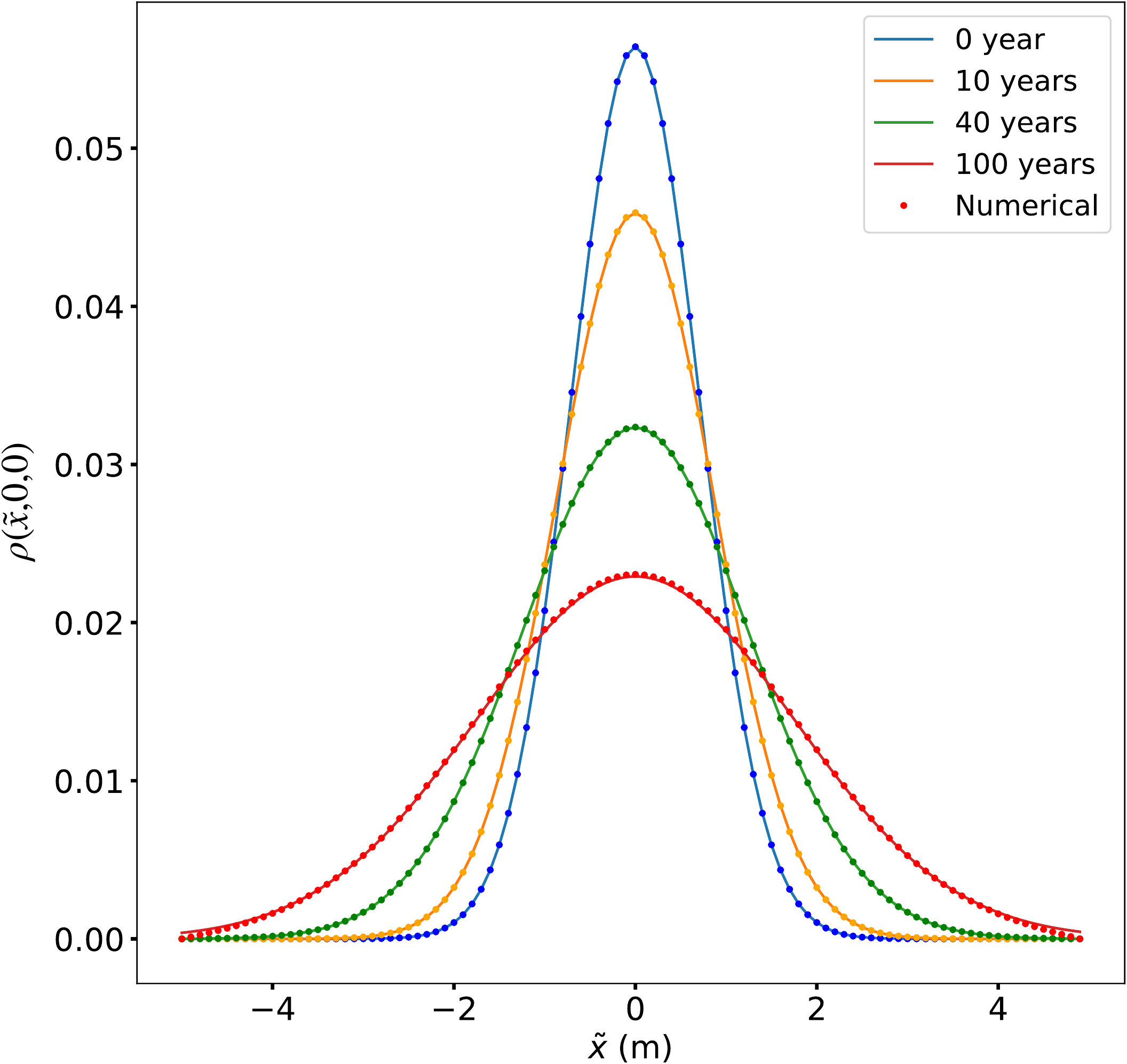}%
\label{fig1b:passivecompaz}}   
        \caption{The probability density function computed from Eqs. \eqref{eq:FPE} and \eqref{eq:solution_fpe} at different times, for a system with one active Ornstein-Uhlenbeck particle under an external temperature gradient pointing in the negative $\tilde{z}$-direction.  The analytic solution from Eq. \eqref{eq:solution_fpe} is given by the solid lines and the numerical solution of Eq. \eqref{eq:FPE} using a finite difference method is given by the dots. At $\tilde{t}=0$, the initial position of the active particle is $\tilde{z}_0=60$m, and the corresponding probability density function is given by Eq. \eqref{eq:solution_fpe_t0}. The evolution of the probability density function 
        (a) Along the $\tilde{z}$-axis at $\tilde{x}=\tilde{y}=0$.  
(b) Along the $\tilde{x}$-axis at $\tilde{y}=0$ and $\tilde{z}=59$m. 
The times are shown in the insets. Both figures are computed for a particle radius $R=10^{-6}$m, with a concentration of impurities $N_i = 100 \mu$M$ $m$^{-2}$, a temperature gradient $|\nabla T| = 0.1$ K$ $m$^{-1}$ and an active diffusivity $\tilde{D}_a=1000 \tilde{D}$.}
        \label{fig1:passivecompaz}
\end{figure}

We rewrite Eq. \eqref{eq:effective_fpe} as
\begin{align}
\frac{\partial}{\partial \tilde{t}}\rho + \left[\tilde{v}-2\frac{\partial}{\partial \tilde{z}}\tilde{D}  \right]\frac{\partial}{\partial \tilde{z}}\rho &= -\rho\left( \frac{\partial}{\partial \tilde{z}}\tilde{v} +\frac{\partial^2}{\partial \tilde{z}^2}\tilde{D} \right) \nonumber\\
&+\tilde{D}_{\text{eff}}\nabla^2\rho \,,
\label{eq:FPE}
\end{align}
where $\tilde{D}_{\text{eff}}(\tilde{z})=\tilde{D}_a+\tilde{D}(\tilde{z})$ is the effective diffusion coefficient.
As was done in \cite{marath-sm-2020}, we define the P\' eclet function as 
\begin{equation}
\text{Pe}(\tilde{z})=\frac{\left[\tilde{v}-2\frac{\partial}{\partial \tilde{z}}\tilde{D}\right]}{\tilde{D}_{\text{eff}}}L\,,
\label{eq:peclet_lim}
\end{equation}
which facilitates the solution of Eq. \eqref{eq:FPE}.  
As a reminder, $L$ is the macroscopic thermal diffusion length scale along the $\tilde{z}$-direction. When $\text{Pe}$ is large, diffusion in the $\tilde{z}$-direction can be neglected relative to regelation driven advection \cite{marath-sm-2020} and Eq. \eqref{eq:FPE}  becomes
\begin{equation}
\frac{\partial}{\partial \tilde{t}}\rho + \tilde{v}\frac{\partial}{\partial \tilde{z}}\rho=-\rho\frac{\partial}{\partial \tilde{z}}\tilde{v} + \tilde{D}_{\text{eff}}\left[\frac{\partial^2}{\partial \tilde{x}^2}\rho+\frac{\partial^2}{\partial \tilde{y}^2}\rho  \right]\,,
\label{eq:FPE_pe_limit}
\end{equation}
the solution to which is given by 
\begin{align}
\rho(\tilde{\bm{r}},\tilde{t})&=\frac{\tilde{z}^3}{(\tilde{z}')^{3/4}}\exp\left(-\frac{\left[(\tilde{z}')^{1/4}-\tilde{z}_0\right]^2}{20+4\tilde{D}_a\tilde{t}} \right) \nonumber\\
&\times \exp\left[ -\frac{(\tilde{x}^2+\tilde{y}^2)}{\left( 1+4\frac{\tilde{D}(\tilde{z})}{\tilde{v}(\tilde{z})}\left[(\tilde{z}')^{1/4}-\tilde{z}\right]+4\tilde{D}_a\tilde{t}\right)} \right]\nonumber\\
&\times \frac{1}{2\pi\sqrt{5\pi}\left(1+4\frac{\tilde{D}(\tilde{z})}{\tilde{v}(\tilde{z})}\left[(\tilde{z}')^{1/4}-\tilde{z}\right]+4\tilde{D}_a\tilde{t}  \right)}\,,
\label{eq:solution_fpe}
\end{align}
with $\tilde{z}'=4\frac{A_3}{A_2^3}\tilde{t}+\tilde{z}^4$. Equation \eqref{eq:solution_fpe} is a consequence of the particular parameters of interest here which insure that the P\' eclet function in Eq. \eqref{eq:peclet_lim} is large so that, as noted above, thermal regelation dominates particle dynamics.  
Clearly, however, this depends on the temperature gradient, particle size and impurity concentration. 
Here for $R=10^{-6}$ m
the regelation induced effective diffusivity is of order $10^{-12}$m$^{2} $s$^{-1}$, whereas for similar sized active particles 
as the algae \textit{Chlamydomonas reinhardtii} \cite{fragkopoulos-arxiv-2021,brun-jcp-2019,leptos-prl-2009}, the bacteria \textit{Escherichia Coli} \cite{weber-fp-2019,kim-book-1996} and \textit{Pseudomonas viscosa} \cite{kim-book-1996}, active droplets \cite{jin-pre-2019,jin-pnas-2017}, and recent speed measurements of biota in ice  \cite{olsen-sr-2019,obertegger-book-2020}, we find a range of active diffusivity; $\tilde{D}_a \in [10\tilde{D}, 2000 \tilde{D}]$. 
\begin{figure}
\centering
\includegraphics[width=1.0\columnwidth]{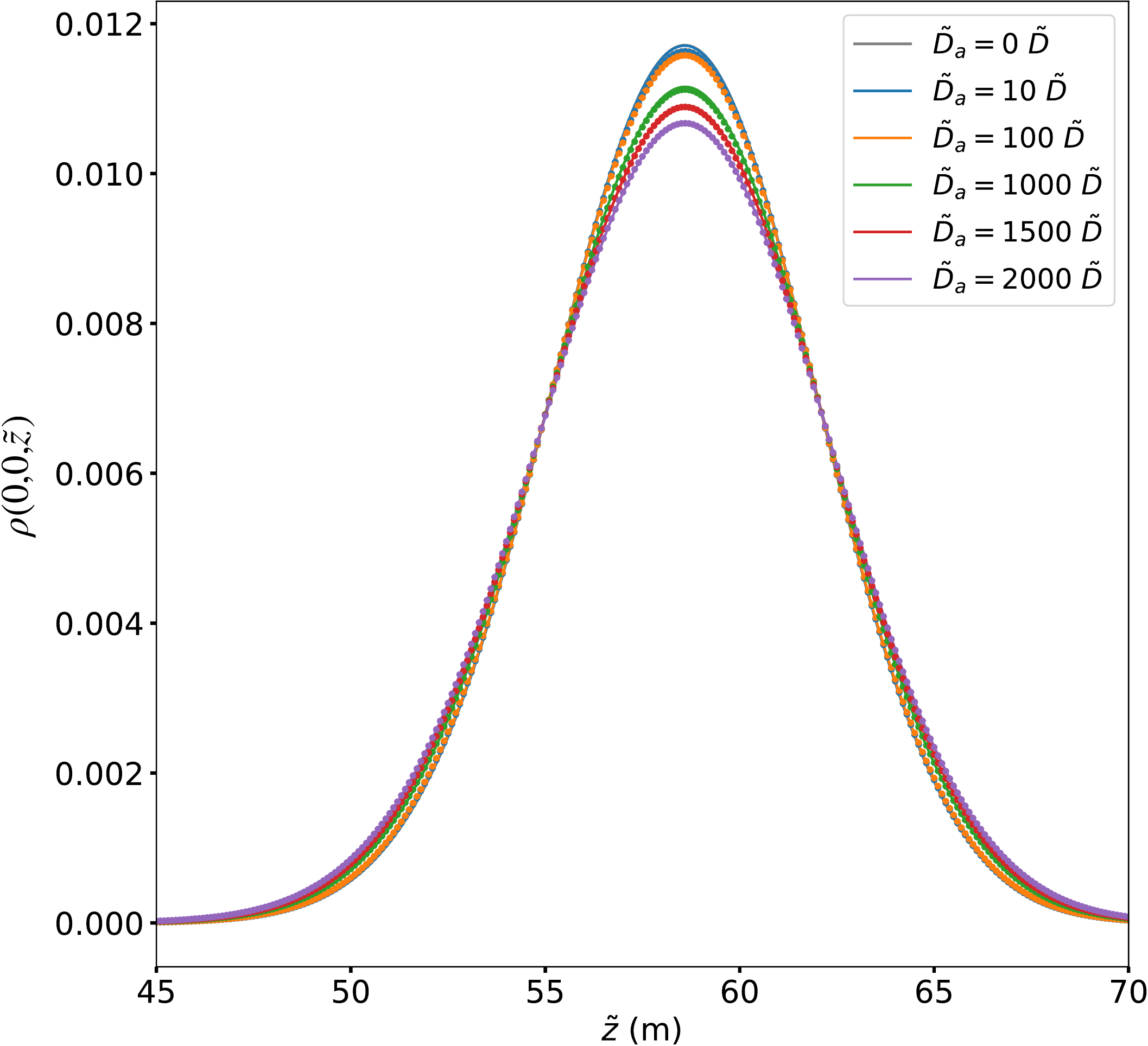}
\caption{The effect of the activity on the dynamics. Evolution of the probability density along the $\tilde{z}$-axis, computed from Eq. \eqref{eq:solution_fpe}, for $\tilde{x}=\tilde{y}=0$, initial condition $\tilde{z}_0 = 60$m at time $\tilde{t}=40$ years for different active diffusivities $\tilde{D}_a$ shown in the inset.  The analytic solution (solid lines), Eq. \eqref{eq:solution_fpe} is compared with numerical solution (dots), computed from Eq. \eqref{eq:FPE}, with $R=10^{-6}$m,  $N_i = 100 \mu$M$ $m$^{-2}$ and $|\nabla T| = 0.1$ K$ $m$^{-1}$.}
\label{fig3:activecompa}
\end{figure}

Figure \ref{fig1:passivecompaz} shows the numerical solution of Eq. \eqref{eq:FPE} and the analytical solution given by Eq. \eqref{eq:solution_fpe}, for an active diffusivity $\tilde{D}_a = 1000 \tilde{D}$.
The initial position of the particle is $\tilde{z}_0=60$m and the corresponding probability density function is given by
\begin{equation}
\rho(\tilde{\bm{r}},\tilde{t}=0) = \frac{1}{2\pi\sqrt{5\pi}}\exp\left[-\frac{(\tilde{z}-\tilde{z}_0)^2}{20}-(\tilde{x}^2+\tilde{y}^2)\right]\,.
\label{eq:solution_fpe_t0}
\end{equation}
Figure \ref{fig1a:passivecompaz}) shows the dependence of $\rho(\tilde{\bm{r}},\tilde{t})$ on the position $\tilde{z}$ parallel to the temperature gradient, 
at $\tilde{x}=\tilde{y}=0$ at different times. We observe that $\rho(\tilde{\bm{r}},\tilde{t})$ spreads and displaces in the direction of higher temperatures, clearly showing the dominant influence of thermal regelation.  
Figure \ref{fig1b:passivecompaz}) shows the dependence of $\rho(\tilde{\bm{r}},\tilde{t})$ on the position $\tilde{x}$ for $\tilde{y}=0$ and $\tilde{z}=59$m at different times.
Clearly the behavior parallel to the temperature gradient is both advective and diffusive, whereas that perpendicular is diffusive, showing the competition inherent to this system.

To assess the importance of activity on the dynamics we vary the active diffusivity from low to high viz., $\tilde{D}_a \in [10\tilde{D}, 2000 \tilde{D}]$.
Figure \ref{fig3:activecompa} shows the evolution of $\rho(\tilde{\bm{r}},\tilde{t} = 40 \text{yr})$, along the $\tilde{z}$-axis, for the same particle size, impurity concentration and temperature gradient as in Fig. \ref{fig1:passivecompaz} with $\tilde{x}=\tilde{y}=0$.  
The active diffusivity $\tilde{D}_a$ affects the effective diffusion coefficient $\tilde{D}_{\text{eff}}$. When the activity is neglected, we recover the results from \cite{marath-sm-2020}. However, as the activity increases, the effective diffusion coefficient also increases. In the case when the activity becomes extremely large, the P\' eclet function decreases substantially and the limit of large P\' eclet is no longer valid. However, this behavior is unrealistic for the parameters of interest \cite{brun-jcp-2019,leptos-prl-2009,weber-fp-2019,kim-book-1996,olsen-sr-2019,obertegger-book-2020}. Therefore, although the effective diffusion coefficient increases, the large P\' eclet limit is still valid, as shown in Fig. \ref{fig3:activecompa}. The analytic solution (solid lines), Eq. \eqref{eq:solution_fpe}, compares well with numerical solutions (dots), Eq. \eqref{eq:FPE}. However, the increase in the diffusivity induced by the activity changes the precision of the dating method, creating uncertainties that should be of interest \cite{rupper-jg-2015,royer-springer-1983,augustin-nature-2004}.

\begin{figure}
\centering
\includegraphics[width=1.0\columnwidth]{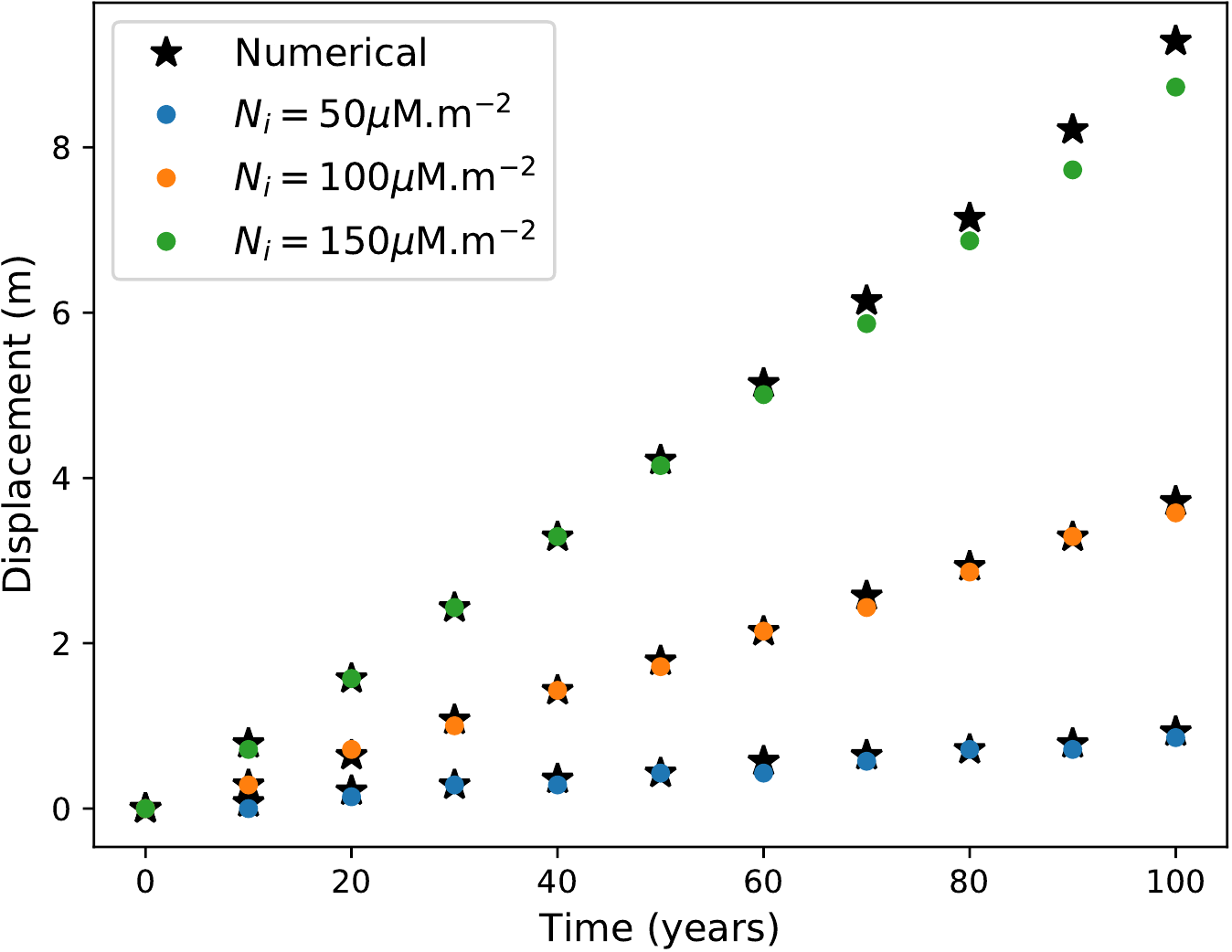}
\caption{The displacement dynamics along the $\tilde{z}$-axis, for different concentration of impurities $N_i$. Here,  the particle radius is $R=10^{-6}$m and initial the condition is $\tilde{z}_0 = 60$m, with a temperature gradient $|\nabla T| = 0.1$ K$\cdot$m$^{-1}$ and an active diffusivity $\tilde{D}_a=1000 \tilde{D}$.}
\label{fig4:activecompa}
\end{figure}

To gauge the effect of impurities on the dynamics, we vary $N_i$, from $50\mu$Mm$^{-2}$ to $150\mu$Mm$^{-2}$ and compute the evolution of the displacement along the $\tilde{z}$-axis, for an active diffusivity of $\tilde{D}_a = 1000 \tilde{D}$. The displacement is calculated by taking the difference between the first moments of the probability density function Eq. \eqref{eq:solution_fpe} and its initial condition Eq. \eqref{eq:solution_fpe_t0}.  The results shown in Fig. \ref{fig4:activecompa} are in agreement with a previous work \cite{marath-sm-2020}. Due to the sensitivity of the premelted film thickness with impurity concentration, the flux of unfrozen water increases with concentration and hence so too does the regelation rate, as shown in Fig. \ref{fig4:activecompa}.
The sensitivity of the dynamics to impurities is reflected in the small discrepancy between the analytic and numerical solution for the long times and large concentrations. It is known that the presence of active particles in ice, such as bacteria or algae, and their concentrations reflect the past state of the environment \cite{takeuchi-jes-2011,abramov-review-2021}, suggesting the importance of a quantitative understanding of the role of impurities in controlling the transport mechanisms of such particles.  

\section{\label{sec:conclusion}Conclusion}
Thermal regelation of active particles provides an interesting framework with many applications \cite{fisher-science-2015,zhang-nature-2005}.  
Here, we have treated the dynamics of one active particle experiencing thermal regelation in three dimensions within the framework of an active Ornstein-Uhlenbeck particle. Firstly, we used a multi-scale expansion to derive the relevant Fokker-Planck equation, Eq. \eqref{eq:effective_fpe}.  Secondly, by taking the limit wherein thermal regelation dominated, an associated P\' eclet function given by Eq. \eqref{eq:peclet_lim} is large, which allows one to find an analytic solution, Eq. \eqref{eq:FPE}, to Eq. \eqref{eq:solution_fpe}.  We showed that, in the regimes of relevance to a range of particles in ice and a large range of active diffusivity, $\tilde{D}_a$,  this limit holds, as reflected in the comparison with the numerical solution.  

An important implication concerns the dating methods in ice core paleoclimatology.  In particular, because of the importance of the diffusivity, our finding that this increases with activity may be reflected in dating uncertainties \cite{extier-qsr-2018,rupper-jg-2015}. Finally, of relevance to both ice core dating and extremophiles is the influence impurities on the dynamics as shown in Fig. \ref{fig4:activecompa}. 
By increasing the concentration of impurities the particle displacement increases substantially and hence we expect this effect to be important for both inert and living particle dynamics.  Indeed, an interesting question concerns directed biolocomotion that opposes thermal regelation.   Do biota ``swim'' against the thermo-molecular pressure gradient?  How do these gradients compete with those associated with nutrients within an ice sheet \cite{wadham-naturecom-2019,holland-bio-2019}? 
Finally, as suggested from recent studies of multiple passive particles \cite{worster-jfm-2021, Worster:2021}, 
generalizing these questions to multiple active particles may reveal intriguing biological effects. Specifically, when active particles aggregate interesting long time collective effects might arise, as observed in the passive colloidal assemblies \cite{reichhardt-pre-2006} and the active case will be particularly intriguing to study. The framework described here provides a starting point for generalization to address these and related questions.

\begin{acknowledgments}
We acknowledge N. Marath, M. Su$\tilde{\text{n}}$\'e Simon, A. Bonfils, R. Eichhorn, S. Ravichandran and W. Moon for helpful discussions and advice, and Swedish Research Council grant no. 638-2013-9243 for support.
\end{acknowledgments}

\newpage

\bibliography{biblio.bib}

\end{document}